\documentclass[fleqn,usenatbib]{mnras}

\usepackage{newtxtext,newtxmath}
\usepackage{float}
\usepackage{ulem}
\usepackage[T1]{fontenc}

\DeclareRobustCommand{\VAN}[3]{#2}
\let\VANthebibliography\thebibliography
\def\thebibliography{\DeclareRobustCommand{\VAN}[3]{##3}\VANthebibliography}

\usepackage{graphicx}	
\usepackage{amsmath}	

\title[VSF from CSST]{Cosmological Forecast of the Void Size Function Measurement from the CSST Spectroscopic Survey}

\author[Y. Song et al.]{
Yingxiao Song$^{1,2}$,
Qi Xiong$^{1,2}$,
Yan Gong$^{1,2,3}$\thanks{Email:gongyan@bao.ac.cn},
Furen Deng$^{1,2}$,
Kwan Chuen Chan$^{6,7}$,\newauthor
Xuelei Chen$^{1,2,4,5}$,
Qi Guo$^{1,2}$,
Jiaxin Han$^{8,9,10}$,
Guoliang Li$^{11}$,
Ming Li$^1$,
Yun Liu$^{1,2}$,
Yu Luo$^{11}$,\newauthor
Wenxiang Pei$^{1,2}$,
and Chengliang Wei$^{11}$
\\\\
$^{1}$National Astronomical Observatories, Chinese Academy of Sciences,20A Datun Road, Beijing 100012, China\\
$^{2}$School of Astronomy and Space Sciences, University of Chinese Academy of Sciences(UCAS),Yuquan Road NO.19A Beijing 100049, China\\
$^{3}$Science Center for China's Space Survey Telescope, National Astronomical Observatories, Chinese Academy of Sciences,\\20A Datun Road, Beijing 100101, China\\
$^{4}$Department of Physics, College of Sciences, Northeastern University, Shenyang 110819, China\\
$^{5}$Centre for High Energy Physics, Peking University, Beijing 100871, China\\
$^{6}$School of Physics and Astronomy, Sun Yat-sen University, 2 Daxue Road, Tangjia, Zhuhai, 519082, China\\
$^{7}$CSST Science Center for the Guangdong-Hongkong-Macau Greater Bay Area, SYSU, Zhuhai, 519082, China\\
$^{8}$Department of Astronomy, School of Physics and Astronomy, Shanghai Jiao Tong University, Shanghai, 200240, People's Republic of China\\
$^{9}$Key Laboratory for Particle Astrophysics and Cosmology (MOE), Shanghai 200240, China\\
$^{10}$Shanghai Key Laboratory for Particle Physics and Cosmology, Shanghai 200240, China\\
$^{11}$Purple Mountain Observatory, Chinese Academy of Sciences, Nanjing, 210023, PR China
}
\date{Accepted XXX. Received YYY; in original form ZZZ}

\pubyear{2015}

\begin{document}
\label{firstpage}
\pagerange{\pageref{firstpage}--\pageref{lastpage}}
\maketitle

\begin{abstract}

Void size function (VSF) contains the information of the cosmic large-scale structure (LSS), and can be used to derive the properties of dark energy and dark matter. We predict the VSFs measured from the spectroscopic galaxy survey operated by China's Space Survey Telescope (CSST), and study the strength of cosmological constraint. We employ a high-resolution Jiutian simulation to get CSST galaxy mock samples based on an improved semi-analytical model. We identify voids from this galaxy catalog using the watershed algorithm without assuming a spherical shape, and estimate the VSFs at different redshift bins from $z=0.5$ to 1.1. We propose a void selection method based on the ellipticity, and assume the void linear underdensity threshold $\delta_{\rm v}$ in the theoretical model is redshift-dependent and set it as a free parameter in each redshift bin. The Markov Chain Monte Carlo (MCMC) method is adopted to implement the constraints on the cosmological and void parameters.  We find that the CSST VSF measurement can constrain the cosmological parameters to a few percent level. The best-fit values of $\delta_{\rm v}$ are ranging from $\sim-0.4$ to $-0.1$ as the redshift increases from 0.5 to 1.1, which has a distinct difference from the theoretical calculation with $\delta_{\rm v}\simeq-2.7$ assuming the spherical evolution and using particles as tracer. Our method can provide a good reference for the void identification and selection in the VSF analysis of the spectroscopic galaxy surveys.
 \end{abstract}




\section{Introduction}

In the past decades, thanks to high-precision cosmological observations and powerful simulations, the study of cosmic large-scale structure (LSS) has made significant progress. Cosmic void, also known as the low-density regions in the Universe, has begun to show its great potential, and is becoming an effective and competitive probe in cosmological research \citep{pisani2019cosmic,moresco2022unveiling,2023arXiv231211241S,2023JCAP...05..031S}. 
Since cosmic void has the characteristics of large volume, low density and linear evolution, it is suitable to extract information of the evolution of cosmic LSS, and explore the structure growth rate, nature of dark energy and modified gravity theories \citep[e.g.][]{cai2015testing,pisani2015counting,zivick2015using,pollina2016cosmic,achitouv2016testing,sahlen2016cluster,falck2018using,sahlen2018cluster,paillas2019santiago,perico2019cosmic,verza2019void,contarini2021cosmic,mauland2023void}.  

Cosmic voids are sensitive to several cosmological geometrical and dynamical effects, like Alcock-Paczy\'{n}ski effect \citep[e.g.][]{sutter2012first,sutter2014measurement,hamaus2016constraints,mao2017cosmic,correa2021redshift,2022A&A...658A..20H}, baryonic acoustic oscillations (BAO) \citep[e.g.][]{kitaura2016signatures,liang2016measuring,chan2021volume,forero2022cosmic,khoraminezhad2022cosmic}, and redshift space distortions (RSD) \citep[e.g.][]{paz2013clues,cai2016redshift,2017JCAP...07..014H,chuang2017linear,nadathur2019accurate,nadathur2020completed,correa2022redshift}.
Various current and upcoming galaxy spectroscopic surveys can be or have been used to study cosmic voids, such as Baryon Oscillation Spectroscopic Survey  \citep[BOSS,][]{2017MNRAS.470.2617A} and Extended Baryon Oscillation Spectroscopic Survey  \citep[eBOSS,][]{2016AJ....151...44D} from the Sloan Digital Sky Survey  \citep[SDSS,][]{2017AJ....154...28B}, Dark Energy Survey  \citep[DES,][]{2016MNRAS.460.1270D}, and Dark Energy Spectroscopic Instrument  \citep[DESI,][]{2016arXiv161100036D}.

Besides, the next-generation spectroscopic surveys are capable to map deeper Universe in higher precision, and hundred millions of galaxy spectra will be obtained, e.g. MegaMapper \citep{2019BAAS...51g.229S}, WideField Spectroscopic Telescope \citep[WST,][]{2019BAAS...51g..45E}, the MaunaKea Spectroscopic Explorer \citep[MSE,][]{2019arXiv190303158P}, MUltiplexed Survey Telescope\footnote{\url{https://must.astro.tsinghua.edu.cn/en}} (MUST), etc. In addition, the slitless spectroscopic observations carried out by Stage-IV space-borne surveys would also provide valuable and enormous information about the cosmic LSS, such as {\it Euclid} \citep{2022A&A...662A.112E}, Nancy Grace Roman Space Telescope (RST), and China's Space Survey Telescope (CSST) \citep{zhan11,zhan2021csst,gong,2023MNRAS.519.1132M}. These surveys undoubtedly can increase the cosmic void samples dramatically, and in the mean time, make higher requirement for the data analysis of accurately extracting information.

In this work, we study the cosmological constraints of the measurements of void size function (VSF) by the CSST spectroscopic galaxy survey. The VSF denotes the number density of voids as a function of size at a given redshift, which can reflect the properties of the LSS. The theoretical model of the VSF is constructed by studying the hierarchical evolution of the void population within the framework of excursion-set theory \citep{sheth2004hierarchy} and later extended to the volume conserving model by \cite{jennings2013abundance}. The void size function has been widely studied and proven to be an effective cosmological probe, it has been used in galaxy spectroscopic surveys and cosmological simulations to constrain cosmology \citep{pisani2015counting,contarini2021cosmic,2022A&A...667A.162C,contarini2023cosmological,2023MNRAS.522..152P,2024arXiv240114451V}.

The CSST is a 2m space-based telescope. It will cover a sky area of 17500 deg$^2$ with both multi-band photometric imaging and slitless grating spectroscopic surveys in about ten years. It has seven photometric bands (i.e. $NUV$, $u$, $g$, $r$, $i$, $z$ and $y$) and three spectroscopic bands (i.e. $GU$, $GV$ and $GI$), covering wavelength range 250-1000 nm. The CSST spectroscopic survey can reach a magnitude limit $\sim$23 AB mag for 5$\sigma$ point source detection. The angular resolution is $\sim0.3''$ within 80\% energy concentration radius for the spectroscopic survey, and the spectral resolution $R = \lambda/\Delta \lambda$ is better than 200. More than one hundred million galaxy spectra are expected to be measured at $0<z<2$. So the CSST spectroscopic survey would be a powerful observation for studying the evolution of the LSS.

We first generate the mock galaxy catalog using simulations and considering the CSST instrumental design and strategy of the CSST spectroscopic survey. Then we identify voids and create the void mock catalog adopting a method based on Voronoi tessellation and watershed algorithm, and derive the  information of void volume-weighted center, effective radius, ellipticity, etc. The shapes of these voids are relatively arbitrary without assuming a simple spherical shape. In order to obtain an accurate result, we also select a `high-quality' void sample for the VSF analysis according to the void ellipticity. After computing the theoretical VSF based on the halo model, we compare the mock observational VSFs with theoretical ones at different redshift bins from $z=0.5$ to 1.1, and the Markov Chain Monte Carlo (MCMC) method is adopted to perform the constraints. We constrain the cosmological parameters, such as the total matter density parameter $\Omega_\text{m}$ and dark energy equation of state $w$,  and the underdensity thresholds for void formation $\delta_\text{v}$ by using the VSF mock data at different redshifts.

The paper is organized as follows: In Section \ref{sec:data}, we introduce the simulation we use, the methods of generating the mock galaxy and void catalogs of the CSST spectroscopic galaxy surveys; In Section \ref{sec:vsf}, we discuss the calculation and estimation of the theoretical and mock observational VSFs; In Section \ref{sec:mcmc}, we predict the constraints on relevant cosmological parameters and void parameters; We summarize our work in Section \ref{sec:conclusion}.

\section{Mock Catalogs} \label{sec:data}

\subsection{Simulation} \label{sec:sim}

We employ the high-resolution N-body simulations from Jiutian simulation to study the expected void size function. The cosmological parameters are from $\it Planck$2018 \citep{2020A&A...641A...6P}, with the density parameters of total matter $\Omega_\text{m} = 0.3111$, baryon $\Omega_\text{b} = 0.0490$, and dark energy $\Omega_\Lambda = 0.6899$, the amplitude of matter fluctuation $\sigma_8 = 0.8102$, spectral index $n_\text{s} = 0.9665$, and the dimensionless Hubble constant $h = 0.6766$. The simulation is carried out with $6144^3$ particles in a box of 1 $h^{-1}$Gpc per side using the GADGET-3 code \citep{2001NewA....6...79S,2005MNRAS.364.1105S}, and the mass of each particle is about $3.73 \times 10^8$ $h^{-1}M_\odot$. The simulation begins at initial redshift $z_i = 127$ with 128 snapshots outputting between $z_i$ and $z = 0$. The friend-of-friend and subfind algorithm are used to identify the dark matter halos and substructures \citep{2001NewA....6...79S,2005MNRAS.364.1105S}.
Since the number of high-redshift galaxies that can be detected by the CSST spectroscopic survey is limited \citep{gong}, we only consider sources at $z < 1.5$ in our void size function discussion. Specifically, we construct six simulation cubes with the central redshift $z_{\rm c} = [0.3,0.5,0.7,0.9,1.1,1.3]$ to build our galaxy and void catalogs.

In order to account for the RSD and structure evolution effects, we construct each simulation cube by a few slices based on the outputting snapshots at different redshifts in the redshift range of the simulation box. Firstly, we choose the line-of-sight (LOS) direction which is parallel to a box edge, and then splice the slice-like halo catalogs form the corresponding snapshots together, according to their comoving distances. The lower and upper boundaries of a slice are thus the comoving distance $\chi_{\rm l}^{i}$ of the $i$th snapshot at $z^{i}$ and $\chi_{\rm u}^{i+1}$ of the $(i+1)$th snapshot at  $z^{i+1}$. The number of slices for the six simulation cubes are 22, 19, 17, 16, 16, and 15, respectively. Note that these slices are spliced together without any interpolation performance for simplicity, and we use the halo IDs to avoid repeated halos at interfaces of the adjoining snapshots.

\subsection{Galaxy mock catalog} \label{sec:gcat}

The mock galaxy catalog is built by implementing an improved Semi-Analytic Model \citep{henriques2015galaxy}. The database contains galaxy emission line luminosity produced by post-processing as described in \cite{2024MNRAS.529.4958P}, which can be used to precisely measure the galaxy redshift, and select galaxies that can be detected by the CSST spectroscopic survey. We choose four strong emission lines, i.e. H$\alpha$, H$\beta$, [OIII] and [OII], as indicators to estimate the signal-to-noise ratio (SNR), and then decide whether a galaxy can be detected according to the detection threshold. For simplicity, we treat each galaxy as a point source, since the region of emitting lines is usually small compared to the full size of an emission-line galaxy. 

Following \citet{cao2018testing} and \citet{2022MNRAS.515.5894D}, the SNR per spectral resolution unit for the spectroscopic sample can be estimated by
\begin{equation}
    \text{SNR}=\frac{C_\text{s} t_\text{exp}\sqrt{N_\text{exp}}}{\sqrt{C_\text{s} t_\text{exp}+N_\text{pix}[(B_\text{sky}+B_\text{det})t_\text{exp}+R_\text{n}^2]}}\label{eq1},
\end{equation}
where $t_\text{exp}=150\,\rm s$ is the exposure time, $N_\text{exp} = 4$ is the number of exposures \citep{gong},  $N_\text{pix}=\Delta A/l_\text{p}^2$ is the number of detector pixels covered by an object. Here $\Delta A$ is the pixel area on the detector, assumed to be the same for all galaxies for simplicity, and $l_\text{p} = 0.074''$ is the pixel size. The point-spread function (PSF) is assumed to be a 2D Gaussian distribution with the radius of 80\% energy concentration $\sim0.3''$ in the CSST spectroscopic survey. $B_\text{det}$ = 0.02 $e^-\text{s}^{-1}\text{pixel}^{-1}$ is the dark current of the detector, and $R_\text{n}$ = 5 $e^-\text{s}^{-1}\text{pixel}^{-1}$ is the read noise. $C_\text{s}$ is the counting rate from galaxy, and for emission line $i$ at frequency $\nu_i$, we have
\begin{equation}
    C_\text{s}^i = A_\text{eff}T_X\left(\frac{\nu_i}{z+1}\right)\frac{F_\text{line}^{i}}{h\nu_i/(z+1)},\label{eq2}
\end{equation}
where $A_\text{eff} = 3.14$ m$^2$ is the CSST effective aperture area, and $T_X$ is the total throughput for band $X$ including filter intrinsic transmission, detector quantum efficiency, and mirror efficiency. The redshift $z$ is involving the cosmological redshift $z_\text{cos}$ and peculiar motion redshift $z_\text{pec}$. We have $1+z = (1+z_\text{cos})(1+z_\text{pec}) = (1+z_\text{cos})(1+v_{\rm pec}/c)$, where $v_{\rm pec}$ is the LOS component of peculiar velocity, and we assume a 0.2\% error to each redshift for the accuracy of CSST slitless spectral calibration. $F_\text{line}^{i}$ is the flux of the emission line $i$ which can be obtained from the simulation. $B_\text{sky}$ in Equation~(\ref{eq1}) is the sky background in $e^-\text{s}^{-1}\text{pixel}^{-1}$, which is given by
\begin{equation}
    B_\text{sky} = A_\text{eff}\int I_\text{sky}(\nu)T_\text{X}(\nu)l_\text{p}^2\frac{d\nu}{h\nu},\label{eq3}
\end{equation}
where $I_\text{sky}$ is the surface brightness of the sky background, including earthshine and zodiacal light \citep{2011acsi.book...11U}. We find that $B_\text{sky}=0.016$, 0.196, and 0.266 $e^-\text{s}^{-1}\text{pixel}^{-1}$ for $GU$, $GV$ and $GI$ bands, respectively. 

We select galaxies if $\rm SNR\ge10$ for any emission line of the four lines H$\alpha$, H$\beta$, [OIII] and [OII] in any spectroscopic band to form the galaxy mock catalog. We find that the number density of galaxies are $\bar n =  1.5\times 10^{-2}, 4.3\times 10^{-3}, 1.2\times 10^{-3}, 4.6\times 10^{-4}, 2.0\times 10^{-4}$, and $9.0\times 10^{-5}$  $h^3{\rm Mpc}^{-3}$ for the six simulation cubes we use from $z=0.3$ to 1.3, respectively. These number densities are consistent with the results for the CSST spectroscopic survey derived from the zCOSMOS catalog \citep{gong}.

\subsection{Void mock catalog}

We identify cosmic voids with Void IDentification and Examination toolkit\footnote{\url{https://bitbucket.org/cosmicvoids/vide\_public/src/master/}} \citep[\texttt{VIDE},][]{vide}, a watershed void finding algorithm based on ZOnes Bordering On Voidness \citep[\texttt{ZOBOV},][]{zobov}. This code first estimates a density field from the tracer distribution using Voronoi tessellation, and then applies a watershed algorithm to find localized density minima. \citep{watershed}. No assumptions are made for the shape of the identified voids by \texttt{VIDE}, so that we can find more low density areas as voids with natural shapes. \texttt{VIDE} can provide void properties such as the volume-weighted center, effective radius and ellipticity, which is quite helpful in void analysis. It has become an important tool in the  studies of large-scale structure of the Universe with applications in various void analyses  \citep{sutter2012first,sutter2014measurement,hamaus2016constraints,2022A&A...667A.162C,contarini2023cosmological}.

\begin{figure}
	\includegraphics[width=\columnwidth]{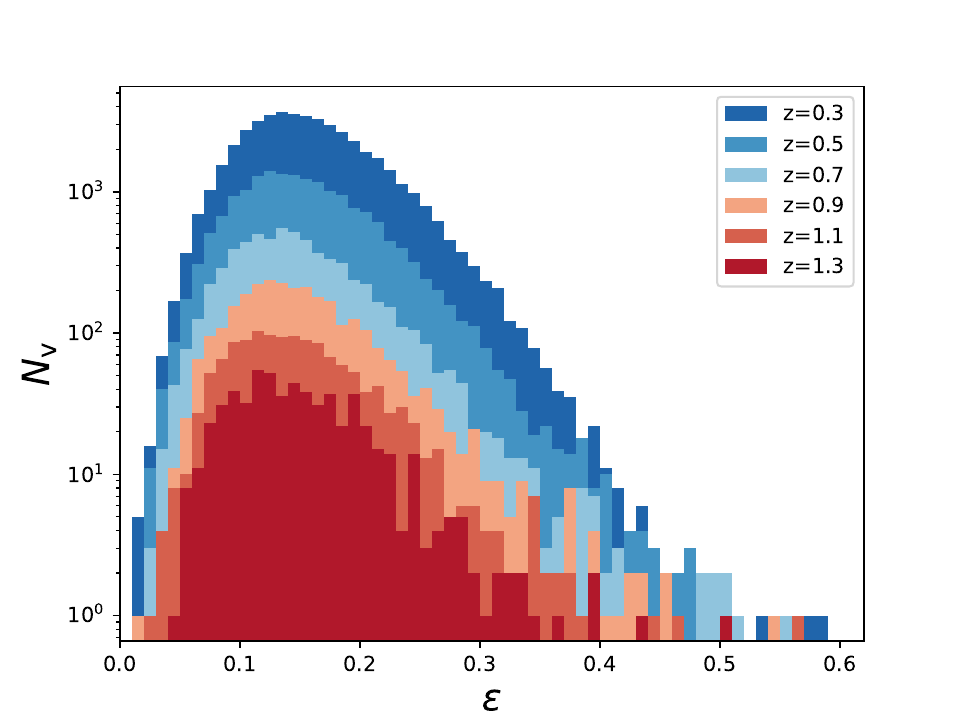}
    \caption{The void ellipticity distribution for different redshift bins. The color from blue to red shows the void ellipticity distribution at redshift bins from $z=0.3$ to 1.3.}
    \label{fig:ved}
\end{figure}

We use the galaxy catalog described in Section~\ref{sec:gcat} to search voids and construct the void mock catalog. 
By making use of \texttt{VIDE}, we find local minima of the density field as the locations of voids, and the void boundaries are determined by the watershed basins of the density field around the minima with no merging of neighbouring zones. As a result, the galaxies in our mock catalog is divided by overdense ridges, and forms separate underdensity areas as voids.
Each void is made up of individual cells obtained by Voronoi tessellation, and each cell contains a particle tracer, i.e. a galaxy. The cell has its own volume $V_ \text{cell}$, and the density of a cell can be estimated by $\rho=1/V_ \text{cell}$.

The radius of a void $R_{\rm v}$ is obtained from an effective sphere, whose volume is the same as the total volume of all cells in a void, and the void effective radius $R_\text{eff}$ is given by
\begin{equation}
R_{\rm v}=R_\text{eff} = (\frac{3V}{4\pi})^{\frac{1}{3}},\label{eq4}
\end{equation}
where $V=\sum V^i_{\rm cell}$ is the void volume.
 The volume-weighted center of the $N$ Voronoi cells $\mathbf{X}_\text{v}$ can be calculated by
\begin{equation}
\mathbf{X}_\text{v} = \frac{1}{V}\sum^N_{i=1}\mathbf{x}_i V^i_{\rm cell},\label{eq5}
\end{equation}
where $\mathbf{x}_i$ is the position of the galaxy in cell $i$ for a given void. We also estimate void shapes by taking account of void member galaxies and constructing the inertia tensor:
\begin{equation}
M_\text{xx} = {\sum_{i=1}^{N}}(y^2_i+z^2_i)
, \ M_\text{xy} = -{\sum_{i=1}^{N}}x_iy_i,\label{eq6}
\end{equation}
where $M_\text{xx}$ and $M_\text{xy}$ are the diagonal and off-diagonal components, and $x_i$,$y_i$, and $z_i$ are the coordinates of the galaxy in cell $i$ relative to the void volume-weighted center. The other components, i.e. $M_{\rm yy}$, $M_{\rm zz}$, $M_{\rm xz}$, and $M_{\rm yz}$, also can be calculated using Equation~(\ref{eq6}). The inertia tensor reflects the distribution of galaxies inside the void, and the galaxy near the edges of the void will be given a higher weight. Then we use the inertia tensor to compute eigenvalues and eigenvectors, and evaluate the ellipticity by
\begin{equation}
\epsilon = 1 - \left(\frac{J_1}{J_3}\right)^{1/4},\label{eq7}
\end{equation}
where $J_1$ and $J_3$ are the smallest and largest eigenvalues of the inertia tensor. For the ellipticity, we have $0 < \epsilon < 1$, and the void shape is flatter when $\epsilon$ becomes larger. We show the void ellipticity distribution for different redshift bins in Figure \ref{fig:ved}. We can see the void ellipticity distributions have a similar shape at different redshift bins with a peak between $\epsilon=0.1$ and 0.2, and most of voids have $\epsilon<0.5$. This indicates that large number of the voids we find have spherical-like shapes.

\begin{table}
	\centering
	\caption{The number densities of galaxies and voids ($>$ 5 $h^{-1}\text{Mpc}$) of our mock catalogs from $z=0.3$ to 1.3. In this sample, the number densities of voids with $\epsilon <0.15$ are also shown.}
	\label{tab:1}
	\begin{tabular}{cccc} 
		\hline
		Redshift & Galaxy & Void & Void ($\epsilon<0.15$)\\&$(h^3\text{Mpc}^{-3})$&$(h^3\text{Mpc}^{-3})$&$(h^3\text{Mpc}^{-3})$ \\
		\hline
        0.3& $1.5 \times 10^{-2}$& $4.8 \times 10^{-5}$&$2.3 \times 10^{-5}$\\
        0.5& $4.3 \times 10^{-3}$& $1.8 \times 10^{-5}$&$9.1 \times 10^{-6}$\\
        0.7& $1.2 \times 10^{-3}$& $6.5 \times 10^{-6}$&$3.6 \times 10^{-6}$ \\
        0.9& $4.6 \times 10^{-4}$& $2.9 \times 10^{-6}$&$1.5 \times 10^{-6}$\\
        1.1&$2.0 \times 10^{-4}$& $1.3 \times 10^{-6}$&$7.3 \times 10^{-7}$\\
        1.3& $9.0 \times 10^{-5}$& $6.0 \times 10^{-7}$&$3.3 \times 10^{-7}$ \\
		\hline
	\end{tabular}
\end{table}

\begin{table}
	\centering
	\caption{The mean, minimum and maximum radii of voids ($>$ 5 $h^{-1}\text{Mpc}$ and $\epsilon <0.15$) from $z=0.3$ to 1.3. The range of void radius used for the VSF analysis we choose is also shown in a given redshift bin.}
	\label{tab:2}
	\begin{tabular}{ccccc} 
		\hline
		Redshift & $R_\text{v}^\text{mean}$ & $R_\text{v}^\text{min}$ & $R_\text{v}^\text{max}$& Radius range\\&($h^{-1}$Mpc)&($h^{-1}$Mpc)&($h^{-1}$Mpc)&($h^{-1}$Mpc)\\
		\hline
        0.3& 11.75& 5.00&44.64&-\\
        0.5& 17.47& 5.00&61.64&(25,35)\\
        0.7& 25.20& 7.08&76.38& (30,50)\\
        0.9& 33.72& 10.95&89.03&(40,60)\\
        1.1&42.10& 17.26&147.17&(50,70)\\
        1.3& 48.57& 20.87&277.83& -\\
		\hline
	\end{tabular}
\end{table}

In Table~\ref{tab:1}, we show the number densities of galaxies and voids with effective radius $R_{\rm v}=R_{\rm eff}>5\ h^{-1}\text{Mpc}$ of our mock catalogs derived from the simulation in different redshift bins. We filter out voids less than $5\ h^{-1}\text{Mpc}$ to avoid the effects of nonlinear evolution. The number densities of voids in this sample with ellipticity less than 0.15 are also shown. In Table~\ref{tab:2}, we show the average, minimum and maximum radii of voids with $R_{\rm v}>5\ h^{-1}\text{Mpc}$ and $\epsilon <0.15$ at different redshift bins. We can find that, as expected, the number density of voids becomes lower and lower from $z=0.3$ to 1.3, which is similar to the trend of the galaxy number density, while the average void radius becomes larger and larger.

In previous VSF studies, the void catalog is usually trimmed by using the internal density contrast of the voids, i.e. the integrated void density contrast profile. The purpose of this method is to find a void that has a spherical substructure, which  can match the volume conserving model. It  has been shown to be an effective way to study the VSF and can obtain accurate cosmological information.\citep{2017A&A...607A..24R,2019MNRAS.488.5075R,verza2019void,2019MNRAS.488.3526C,contarini2021cosmic,2022A&A...667A.162C,contarini2023cosmological,2024A&A...682A..20C}. In this study, as we show in the next sections, the void linear threshold $\delta_{\rm v}$ is treated as a free paramter, which can provide more flexibility of the void shape in the theoretical model. Hence, unlike previous studies, we do not need to construct or find  perfectly spherical voids in the data for matching the model, but only have to select spherical-like voids by the ellipticity. We will study the effect of void density contrast in our future work.

\section{Void Size Function} \label{sec:vsf}

\begin{figure*}
	\includegraphics[scale=0.6]{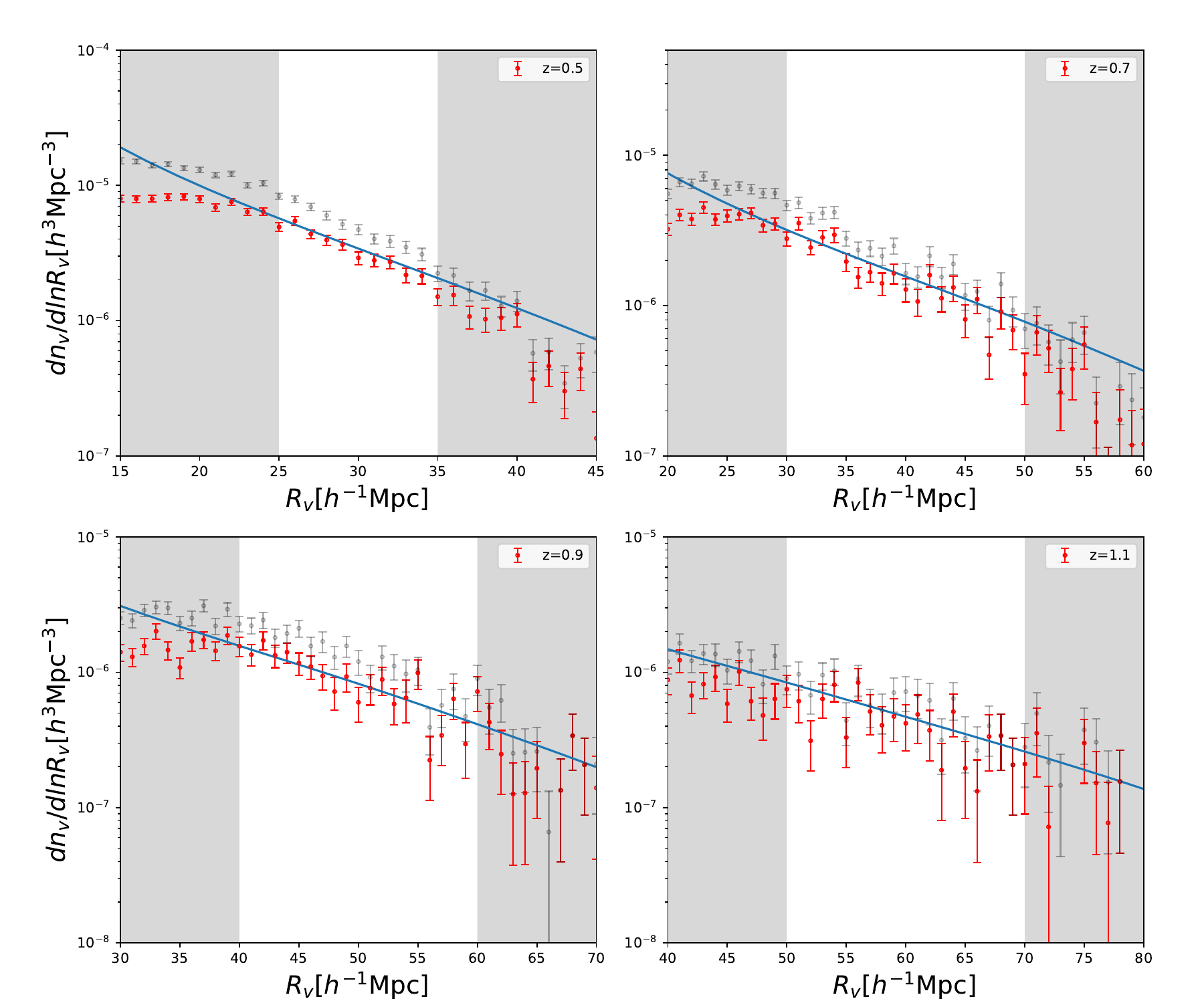}
    \caption{The predicted void size functions for voids with $\epsilon<0.15$ (red) and all voids (gray) for the four redshift bins from $z=0.5$ to 1.1 in the CSST spectroscopic survey. The blue curves are the best-fits of theoretical calculation. The gray regions denote the scales excluded in our analysis using the selection criteria based on the void ellipticity distribution and statistical significance.}
    \label{fig:vsf}
\end{figure*}

The most common theory of void size functions nowadays, which describes the number of voids based on their radius, are based on \cite{sheth2004hierarchy}, and extended by \cite{jennings2013abundance}. This theoretical model adopts the same form as the halo mass function and is based on the development of the excursion-set theory. \citep{press1974formation,peacock1990alternatives,cole1991modeling,bond1991excursion,mo1996analytic}.
To evaluate the VSF, we first derive the void mass function based on the halo mass function. Following \cite{chan2014large}, the void mass function in Lagrangian space is given by
\begin{equation} \label{eq8}
\frac{{\rm d}n_{\rm L}}{{\rm d\,ln}M}=\frac{\bar\rho_\text{m}}{M} \mathcal{F}(\nu,\delta_\text{v},\delta_\text{c})\frac{{\rm d}\nu}{{\rm d\,ln}M}.
\end{equation}
Here $\bar \rho_\text{m}$ is the mean dark matter density, $\delta_\text{c}$ is the critical overdensity barrier, which is about 1.686 in the $\Lambda$CDM model, and $\delta_\text{v}$ is the linear underdensity threshold of the void formation. Assuming spherical evolution with void matter density $\rho_{\rm v}=0.2\bar{\rho}_{\rm m}$, in the Einstein-de Sitter model, the linear threshold $\delta_\text{v}$ is found to be -2.717, and -2.731 in the $\Lambda$CDM model, which have small difference in these two models \citep{jennings2013abundance}. $\nu$ is the peak hight, and it can be written as
\begin{equation}
\nu=\frac{|\delta_\text{v}|}{\sigma_M(z)}. \label{eq9}    
\end{equation} 
Here $\sigma_M(z)=\sigma_MD(z)$, where $D(z)$ is the linear growth factor, and $\sigma_M=\sigma_0(R_{\rm L})$ is the root-mean-squared density fluctuation within Lagrangian size $R_{\rm L}$, where $M=\bar\rho_\text{m}V(R_{\rm L})$ and $V(R_{\rm L})\equiv(4/3)\pi R^3_{\rm L}$. $\sigma_0(R_{\rm L})$ can be expressed by
\begin{equation}
\sigma_0(R_{\rm L})=\int \frac{k^2}{2\pi^2}W^2(kR_{\rm L})P(k)dk. \label{eq10}
\end{equation}
Here $P(k)$ is the matter power spectrum of the density fluctuation field and $W(kR_{\rm L})$ is the spherical top-hat window function for size $R_{\rm L}$. We use \texttt{CAMB} to calculate $P(k)$ and $\sigma_0(R_{\rm L})$ in this work \citep{camb}.

In Equation~(\ref{eq8}), $\mathcal{F}(\nu,\delta_\text{v},\delta_\text{c})$ is the first-crossing distribution, which shows the probability of a random trajectory crossing the barrier $\delta_\text{v}$ for the first time at $\nu$ without crossing $\delta_\text{c}$ for $\nu'>\nu$, and it is written as \citep{sheth2004hierarchy}
\begin{equation}
\mathcal{F}(\nu, \delta_\text{v},\delta_\text{c})=\frac{2\mathcal{D}^2}{\nu^3}\sum^{\infty}_{j=1}j\pi\, {\rm sin}(\mathcal{D}j\pi)\,{\rm exp}(-\frac{j^2\pi^2\mathcal{D}^2}{2\nu^2}),\label{eq11}
\end{equation}
where $\mathcal{D}$ is given by
\begin{equation}
\mathcal{D}=\frac{|\delta_\text{v}|}{\delta_\text{c}+|\delta_\text{v}|}.\label{eq12}
\end{equation}
Besides, $\mathcal{F}(\nu,\delta_\text{v},\delta_\text{c})$ also can be approximated by \citep{sheth2004hierarchy}
\begin{equation}
\mathcal{F}_\text{app}(\nu,\delta_\text{v},\delta_\text{c})=\sqrt{\frac{2}{\pi}}\,{\rm exp}(-\frac{\nu^2}{2})\,{\rm exp}(-\frac{|\delta_\text{v}|}{\delta_\text{c}}\frac{\mathcal{D}^2}{4\nu^2}-2\frac{\mathcal{D}^4}{\nu^4}).\label{eq13}
\end{equation}
So Equation~(\ref{eq11}) or (\ref{eq13}) denote the first-crossing distribution for the double-barrier case, including both $\delta_{\rm v}$ and $\delta_{\rm c}$. For large $M$, the first-crossing distribution will reduce to the one-barrier case $\mathcal{F}(\nu, \delta_\text{v})$ without the second exponential term in Equation~(\ref{eq13}), which is also the case for most voids considered in this work.

\begin{figure*}
	\includegraphics[scale=0.6]{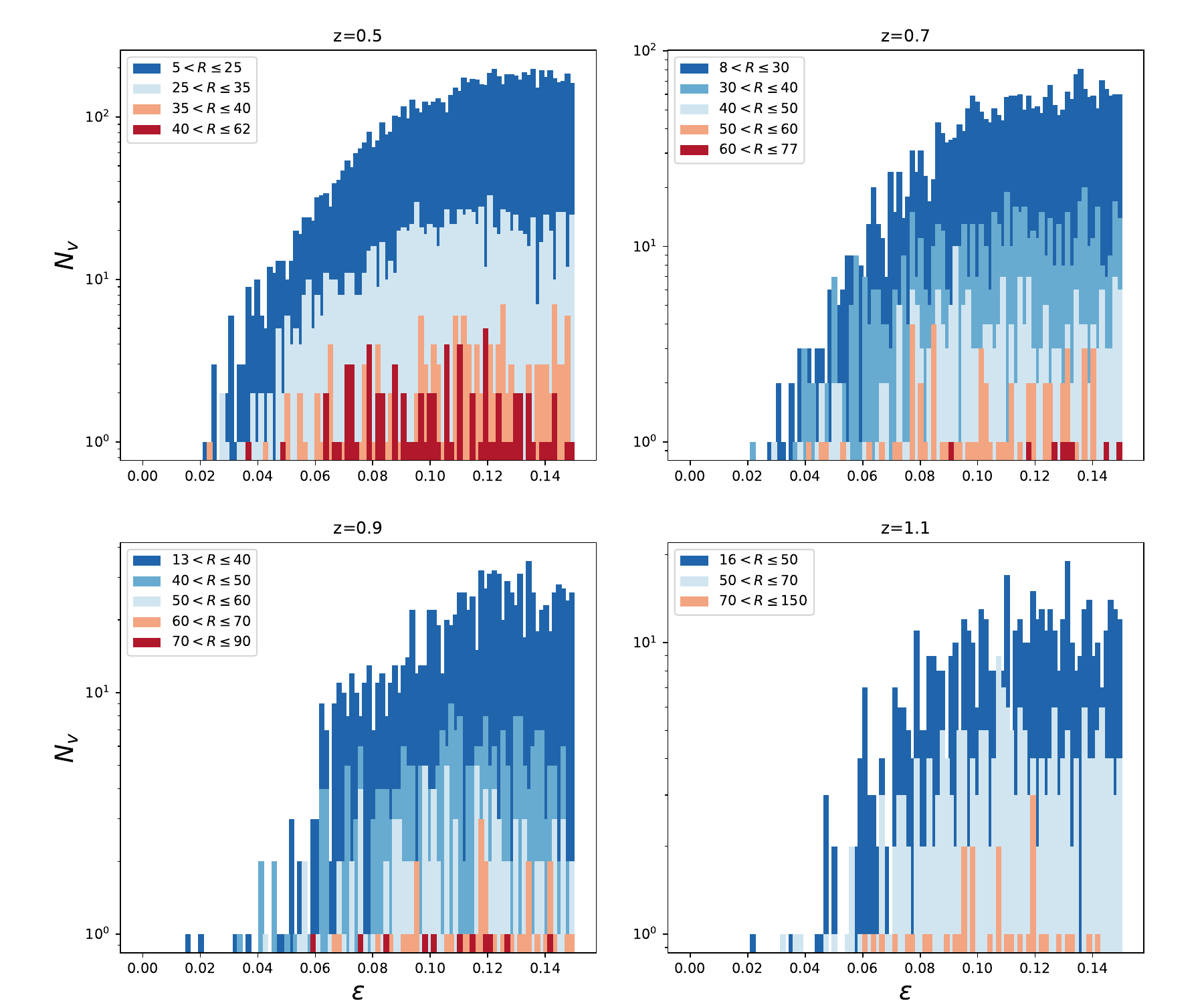}
    \caption{The distributions of void ellipticity with $\epsilon<0.15$ for different void size bins at redshift range from $z=0.5$ to 1.1 in the CSST spectroscopic survey. We can find that most of smaller voids still have relatively large ellipticity, and we will exclude the void size bin with the peak of the ellipticity distribution at $\epsilon>0.12$ and weak statistical significance in the fitting process.}
    \label{fig:epsilon}
\end{figure*}

To map the Lagrangian VSF to the Eulerian one, we need to convert Lagrangian size $R_{\rm L}$ to Eulerian size $R$, and we assume the volume fraction $V{\rm d}n$ conserves \citep{jennings2013abundance}. Then we have
\begin{equation}
\frac{{\rm d}n}{{\rm d\,ln}R}=\frac{V(R_{\rm L})}{V(R)}\frac{{\rm d}n_{\rm L}}{{\rm d\,ln}R_\text{L}}.\label{eq14}
\end{equation}
Hence the theoretical VSF in Eulerian space can be derived based on Equation~(\ref{eq8}) and (\ref{eq14}) for the Eulerian size $R$, which is given by
\begin{equation}
\frac{{\rm d}n}{{\rm d\,ln}R}=(\frac{3}{4\pi R^3})\mathcal{F}(\nu,\delta_\text{v},\delta_\text{c})\frac{d\nu}{{\rm d\,ln}R_\text{L}}.\label{eq15}
\end{equation}

In order to improve the fitting of the void size distribution found by the watershed algorithm in observations, many studies attempt to relax the linear threshold $\delta_\text{v}\simeq-2.7$ or consider it as a free parameter \citep{jennings2013abundance,sutter2014sparse,chan2014large,pisani2015counting,2022A&A...667A.162C,contarini2023cosmological}. However, if we don't set $\delta_\text{v}$ to a fixed value and let it change, the conversion relation between void Eulerian radius $R=R_{\rm v}$ and Lagrangian radius $R_{\rm L}$ needs to be changed according to $\delta_\text{v}$, and then the relation can be well fitted by 
\begin{equation}
R_\text{L}\simeq\frac{R_\text{v}}{(1-\delta_\text{v}/c_{\rm v})^{c_{\rm v}/3}},\label{eq16}
\end{equation}
where $c_{\rm v} =1.594$ \citep{bernardeau1993nonlinear,jennings2013abundance}. In the spherical evolution model, the analytical calculation gives $R_\text{L}=0.58R_\text{v}$ assuming the void matter density $\rho_{\rm v}=0.2\bar{\rho}_{\rm m}$ and $R/R_{\rm L}=(\bar{\rho}_{\rm m}/\rho_{\rm v})^{1/3}$\citep{jennings2013abundance,chan2014large}.  

Since the mock VSF data are generated in the redshift space, which include both the RSD and Alcock-Paczy\'{n}ski effects \citep[AP,][]{1979Natur.281..358A}, the void effective radius of the data in the redshift space $R_{\rm v}^{\rm data}=R_{\rm v}^{\rm (s)}$ is different from the Eulerian radius $R_{\rm v}$ in the theory. Following \cite{correa2021redshift}, we relate $R_{\rm v}^{\rm data}$ and $R_{\rm v}$ by
\begin{equation}\label{eq17}
R_{\rm v}^{\rm data} = R_{\rm v}^{\rm (s)} =  q_{\rm RSD}q_{\rm AP}R_{\rm v},
\end{equation}
where $q_{\rm RSD}$ and $q_{\rm AP}$ denote the correction factor for the RSD and AP effects, respectively. In our void sample, $q_{\rm RSD}$ can be obtained by \citep{correa2021redshift}:
\begin{equation}\label{eq18}
q_{\rm RSD} = 1 - \frac{1}{6}\beta\Delta(R_{\rm v}).
\end{equation}
Here $\Delta(R_{\rm v}) = (1-\delta_{\rm v}/c_{\rm v})^{-c_{\rm v}}-1$ \citep{bernardeau1993nonlinear}, and $\beta = f/b_{\rm g}$, where $f$ is the growth rate and $b_{\rm g}$ is the galaxy bias. We take $\beta$ as free parameters in different redshift bins in our fitting process. The correction for the AP effect is made by scaling factors in the transverse and radial directions \citep{correa2021redshift}:
\begin{equation}\label{eq19}
q_{\rm AP} = \alpha_{\parallel}^{1/3}\alpha_{\perp}^{2/3}.
\end{equation}
Here $\alpha_{\parallel} = H^{\rm fid}(z)/ H(z)$ and $\alpha_{\perp} = D_{\rm A}(z) / D_{\rm A}^{\rm fid}(z)$, and the superscript "fid" indicates that the value is from the fiducial cosmology.

In this work, with the help of Equation~(\ref{eq16}), we set $\delta_{\rm v}$ as a free parameter in a given redshift bin, which means that we assume $\delta_{\rm v}$ is dependent on the redshift, or its evolution cannot be fully described by the linear growth factor $D(z)$. This gives us more flexibility to fit the mock VSF data at different redshifts, and can explore the complexity of the void formation and evolution.

To further reduce the discrepancy between the voids found by \texttt{VIDE} using the watershed algorithm and the theoretical model based on the spherical evolution, and obtain more reliable and accurate fitting results, we select voids according to the ellipticity with $\epsilon<0.15$, i.e. selecting more spherical-like voids. As shown in Figure \ref{fig:ved}, this void ellipticity cut-off is around the peak of the void ellipticity distributions at different redshift bins. By applying this selection criterion, the void number reduction is around 50\% for all redshifts  as indicated in Table~\ref{tab:1}.
In Figure~\ref{fig:vsf}, we show the VSFs for voids with $\epsilon<0.15$ and all voids at different redshift bins from $z=0.5$ to $1.1$. The error bar for each data point is estimated by using the jackknife method. 

In addition, we also analyze the ellipticity distribution with $\epsilon<0.15$ for different void size bins in a given redshift as shown in Figure~\ref{fig:epsilon}. As can be seen, most of smaller voids are still not spherical enough with relatively large ellipticity, and larger voids have small numbers and weak statistics with large fluctuations in distribution. To obtain a better fitting results for the cosmological and void parameters, we only select the void size range with the peak of the ellipticity distribution at $\epsilon\lesssim0.12$ and sufficient statistical significance (SNR>1 for each VSF data point) in the fitting process. 
Hence, the gray regions (small- and large-size parts) in Figure~\ref{fig:vsf} are excluded. Besides, based on these two selection criteria, the first ($z=0.3$) and last ($z=1.3$) redshift bins are also rejected in the fitting process. This is because that all of the void size ranges in the first redshift bins at $z=0.3$ are dominated by voids with $\epsilon\sim0.15$, and the void number densities in the last redshift bins at $z=1.3$ are too low to provide sufficient information for cosmological constraints as shown in Table~\ref{tab:1}. The void size ranges we choose at the four redshift bins from $z=0.5$ to 1.1 are listed in Table~\ref{tab:2}.

\begin{figure*}
	\includegraphics[scale=0.55]{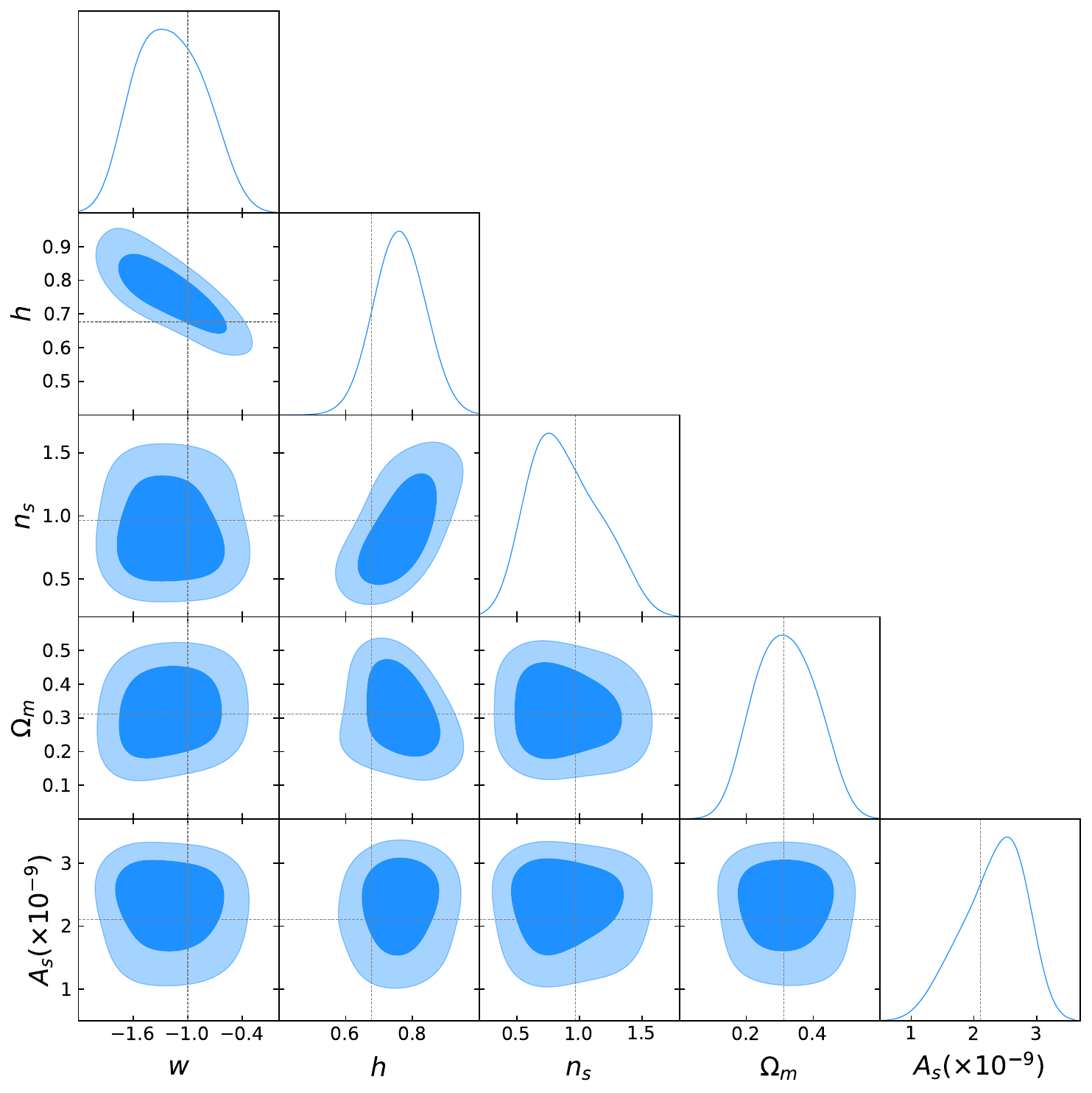}
    \caption{Contour maps of the five cosmological parameters at 68\% and 95\% CL using the VSF mock data at all redshift bins. The 1D PDF (blue curve) for each parameter is also shown. The gray vertical and horizontal dotted lines represent the fiducial values of these parameters.}
    \label{fig:mcmccosmic}
\end{figure*}

\section{Parameter CONSTRAINT} \label{sec:mcmc}

After obtaining the mock data of the VSFs at different redshifts in the CSST spectroscopic survey, we can explore the constraints on the cosmological and void parameters. The $\chi^2$ method is adopted for parameter fitting, which takes the form at $z$ as
\begin{equation}\label{eq20}
\chi_z^2 = \sum\frac{\left[n^{\rm data}_{\rm v}(R_{\rm v},z)-n^{\rm th}_{\rm v}(R_{\rm v},z)\right]^2}{\sigma_\text{v}^2},
\end{equation}
where $n^{\rm data}_{\rm v}(R_{\rm v},z)$ is the VSF mock data at $z$, $n^{\rm th}_{\rm v}(R_{\rm v},z)={\rm d}n/{\rm d\,ln}R$ as shown in Equation (\ref{eq15}) is the  theoretical VSF, and $\sigma_\text{v}$ is the error of the VSF mock data. Then we can calculate the likelihood function as $\mathcal{L}$ $\propto$ exp($-\chi^2$/2). The total chi-square of the void size function can be estimated by $\chi^2_\text{tot} = \sum_{z\,{\rm bins}}\chi^2_z$ for different redshift bins.

We use the \texttt{emcee} package \citep{emcee} to apply Markov Chain Monte Carlo (MCMC) method, which  based on the affine-invariant ensemble sampling algorithm \citep{goodman}. We initialize 112 walkers around the target parameters to get 15000 steps. We then discard the first 10 percent of the steps as burn-in. In Table \ref{tab:3}, we show the free parameters in our model at different redshift, their fiducial values and flat priors. We try to constrain the total matter density parameter $\Omega_\text{m}$, dark energy equation of state $w$, reduced Hubble constant $h$, spectral index $n_{\rm s}$, and amplitude of initial power spectrum $A_{\rm s}$. Since the VSF is not sensitive to baryon density parameter $\Omega_\text{b}$, we do not consider $\Omega_\text{b}$ as a free parameter in the fitting process. The free parameters about void are $\delta_\text{v}^i$ for the four redshift bins from $z=0.5$ to 1.1, which denote the threshold for void formation. And we also set the RSD parameter $\beta$ in the four redshift bins as free parameters. Therefore, we totally have 5 cosmological parameters, 4 void parameters and 4 RSD parameters for the four redshift bins in our model.

\begin{table}
\caption{The fiducial values, prior ranges, best-fit values, errors, and relative accuracies of the five cosmological parameters, the void linear underdensity threshold parameter $\delta_\text{v}^i$ and RSD parameter $\beta^i$ for the four redshift bins from $z=0.5$ to 1.1 have been shown.}
\renewcommand{\arraystretch}{1.5}
\centering
\begin{tabular}{cccc}
\hline\hline
Parameter& Fiducial value& Flat Prior &Best-fit value\\
\hline  
$w$ &-1& (-1.8, -0.2) &$-1.211_{-0.411}^{+0.429}(34.7\%)$\\
$h$ & 0.6766& (0.5, 0.9) & $0.761_{-0.074}^{+0.077} (9.9\%)$\\
$n_\text{s}$ &0.9665& (0.5, 1.5) &$0.837_{-0.248}^{+0.383} (37.7\%)$\\
$\Omega_\text{m}$ & 0.3111& (0.1, 0.5) & $0.315_{-0.097}^{+0.104} (31.9\%)$\\
$A_\text{s}(\times 10^{-9})$ &2.1& (1.0, 3.0) &$2.407_{-0.641}^{+0.428} (22.2\%)$\\
\hline 
$\delta_\text{v}^1$&-& (-2, 0) &$-0.309_{-0.075}^{+0.072} (23.8\%)$\\
$\delta_\text{v}^2$&-& (-2, 0) &$-0.173_{-0.083}^{+0.077} (46.2\%)$\\
$\delta_\text{v}^3$&-& (-2, 0) &$-0.144_{-0.055}^{+0.037} (32.1\%)$\\
$\delta_\text{v}^4$&-& (-2, 0) &$-0.109_{-0.063}^{+0.066} (59.3\%)$\\
\hline 
$\beta^1$&-& (0, 1) &$-0.623_{-0.367}^{+0.259} (50.2\%)$\\
$\beta^2$&-& (0, 1) &$-0.397_{-0.285}^{+0.380} (83.7\%)$\\
$\beta^3$&-& (0, 1) &$-0.619_{-0.387}^{+0.273} (53.4\%)$\\
$\beta^4$&-& (0, 1) &$-0.498_{-0.336}^{+0.340} (68.0\%)$\\
\hline
\end{tabular}
\label{tab:3}
\end{table}

\begin{figure}
	\includegraphics[width=\columnwidth]{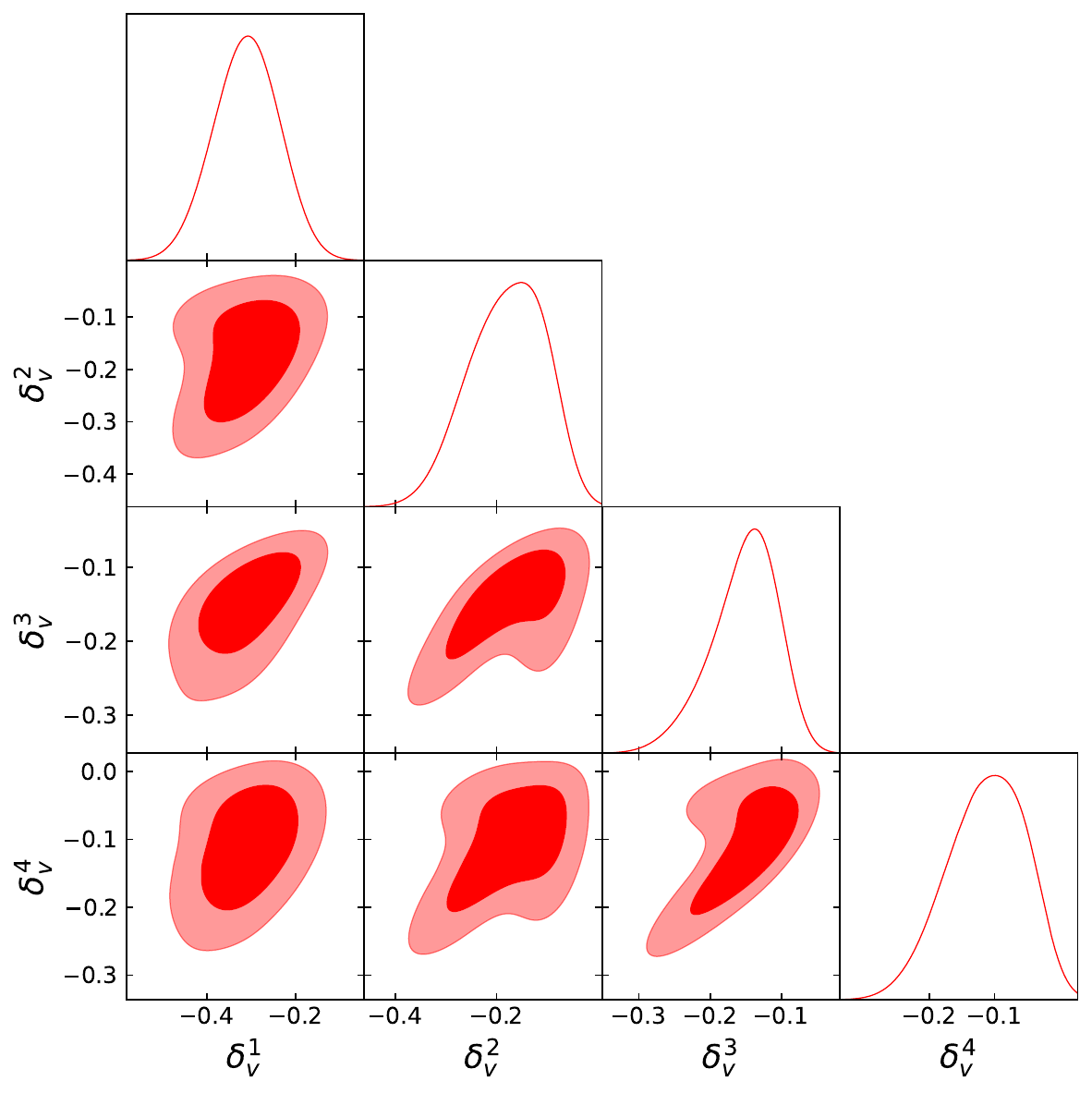}
    \caption{Contour maps of the linear underdensity thresholds for void formation $\delta_\text{v}^i$ at 68\% and 95\% CL using the VSF mock data at all four redshift bins from $z=0.5$ to 1.1. The 1D PDF (red curve) for each parameter is also shown.}
    \label{fig:mcmcdv}
\end{figure}

\begin{figure}
	\includegraphics[width=\columnwidth]{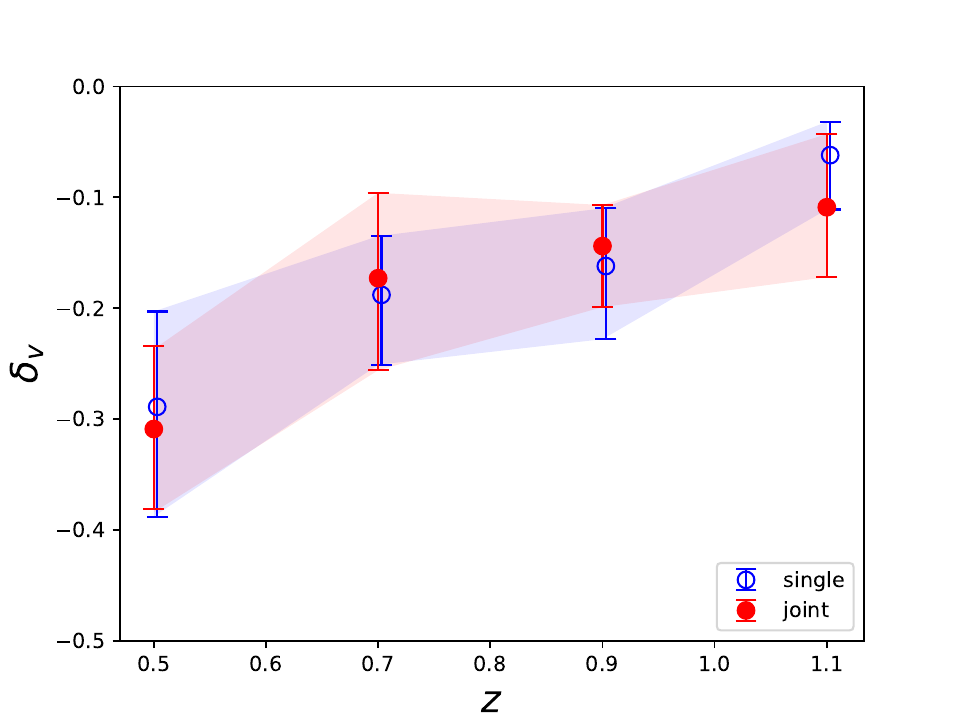}
    \caption{The best-fit values and 1$\sigma$ errors of $\delta_\text{v}^i$ at the four redshift bins. The red and blue data points denote the constraint results using the VSF mock data of all redshift bins and a single redshift bin, respectively.}
    \label{fig:dvcompare}
\end{figure}

In Figure \ref{fig:mcmccosmic} , we show the contour maps at 68\% and 95\% confidence levels (CL) and 1D probability distribution functions (PDFs) of the six cosmological parameters constrained by all of the void size functions from the four redshift bins. The gray dashed line marks the fiducial value of the parameter. We can find that the cosmological parameters we consider can be correctly and well constrained, and the fiducial values are within or close 1$\sigma$ CL of the parameter fitting results. This also proves that the calibration method for the VSF data mentioned in the last section is effective, and could correctly reserve the cosmological information. The details of the best-fit values, 1$\sigma$ errors, and relative accuracies for the six cosmological parameters are shown in Table \ref{tab:3}.

We can see that the current VSF mock data in the CSST spectroscopic survey can obtain  $\sim$30\% and 35\% constraints on $\Omega_\text{m}$ and $w$ in 1$\sigma$ CL, respectively. Comparing to the current cosmological constraint results using the VSFs, e.g. the Baryon Oscillation Spectroscopic Survey (BOSS) DR12 \citep{contarini2023cosmological}, we can achieve comparable constraint power but with more number of cosmological parameters. If considering the full CSST spectroscopic survey with 17,500 deg$^2$, as we estimate, the current constraint accuracies of the cosmological parameters can be improved by almost one order of magnitude, resulting in accuracies of a few percent level.

In Figure~\ref{fig:mcmcdv}, using all the VSF mock data at four redshift bins, the joint constraint results of the linear underdensity thresholds for void formation $\delta_\text{v}^i$ at the four redshift bins are shown in red contours and 1D PDFs. Note that there are no fiducial values for $\delta_\text{v}^i$, since they are not the input parameters in the simulation. The Table \ref{tab:3} shows the details of the best-fit values, 1$\sigma$ errors, and relative accuracies for the four $\delta_\text{v}^i$. 

We notice that the best-fit values of $\delta_\text{v}^i$ are redshift-dependent ranging from about $-0.4$ to $-0.1$ as the redshift increases from $z=0.5$ to 1.1.  This obviously has large discrepancy from the theoretical prediction with a constant $\delta_{\rm v}\simeq-2.7$ assuming the spherical evolution with $\rho_{\rm v}=0.2\bar{\rho}_{\rm m}$ and using particles as tracer  \citep{jennings2013abundance}. This is because that, firstly, the voids in our catalog are identified by the watershed algorithm without considering these assumptions, and secondly, the voids are found in the distribution of galaxies, i.e. biased objects, which are different from the density profile of voids identified by using particles as tracer \citep{2017MNRAS.469..787P,2019MNRAS.487.2836P,2019MNRAS.488.5075R,2019MNRAS.488.3526C}. Besides, as shown in Equation~(\ref{eq16}), this result also implies that the Lagrangian void size $R_{\rm L}$ will be closer and closer to the Eulerian size $R_{\rm v}$ from low to high redshifts, which is consistent with the theoretical expectation.

In order to check this result, we also perform the constraints on the cosmological and void threshold parameters using the VSF mock data in each redshift bin individually. The results are shown in Figure {\ref{fig:dvcompare}}, and the blue and red data points with error bars are the results derived from the data in a single redshift bin and all redshift bins, respectively. We can find that they are basically consistent with each other in $1\sigma$, and conform to the same trend, that both are approaching to zero with increasing redshift. Because there is almost no effect constraint on each cosmological parameter using the data in a single redshift bin, we do not show the constraint result of the cosmological parameters here. This redshift variation of  the linear threshold $\delta_\text{v}$ is also consistent with the result indicated by \citet{2022A&A...667A.162C}, where they use a nonlinear underdensity threshold $\delta_{\rm v}^{\rm NL}$ to study the VSFs measured by {\it Euclid}, which can be converted to the linear threshold $\delta_\text{v}$.
In addition we have also constrained the $\beta^i$ associated with RSD, the results are displayed in Table \ref{tab:3}, and the associated contours and 1D PDFs are shown in Appendix~\ref{sec:appendix}.

\section{Summary} \label{sec:conclusion}

In this work, we study the VSFs measured by the CSST spectroscopic survey at different redshifts, and explore the constraint power for the cosmological and void parameters. We generate the galaxy mock catalog based on Jiutian simulation from $z=0.3$ to 1.3, and consider the instrumental design and survey strategy. We use {\tt VIDE} to identify voids in the galaxy mock catalog by adopting the watershed algorithm, and obtain the void information such as volume-weighted center, effective radius $R_{\rm v}$ and ellipticity $\epsilon$. 

Since the theoretical VSF model is based on the spherical evolution, we also perform a selection using the void ellipticity to find the voids with $\epsilon<0.15$ and $R_{\rm v}>5\,h^{-1}{\rm Mpc}$. The void ellipticity distributions for different void size bins at a given redshift are also explored. In order to obtain a better constraint result, we only use the voids in a size bin with the peak of the ellipticity distribution at $\epsilon\lesssim0.12$ and sufficient statistical significance. As a result, the first ($z=0.3$) and last ($z=1.3$) redshift bins are excluded, and only the voids with middle sizes in each redshift bin from $z=0.5$ to 1.1 are used in the constraint process.

We find that the selected VSF data can correctly reserve the cosmological information and could be well fitted by the theoretical model. The fitting results of the cosmological parameters, such as $\Omega_{\rm m}$ and $w$, are consistent with the fiducial values in 1$\sigma$ CL. We also assume that the linear underdensity threshold $\delta_{\rm v}$ is redshift dependent, and set it as a free parameter in each of the four redshift bins. After the fitting process, we find that $\delta_{\rm v}$ is indeed varying from $\sim-0.4$ to $\sim-0.1$ as the redshift increases from $z=0.5$ to 1.1. This has significant difference from the theoretical calculation with a constant $\delta_{\rm v}\simeq-2.7$ assuming the spherical evolution, since we are using galaxies as biased tracers and the watershed algorithm to identify voids with relatively natural shapes.

We note that the simulation can only cover a small part of survey area of the CSST spectroscopic survey, and the constraint accuracies of the cosmological and void parameters are expected to be improved to a few percent level, if using the data from full 17,500 deg$^2$ CSST wide-field survey area. The method we propose to identify and select voids also can be a reference for the future VSF study in the spectroscopic galaxy surveys.

\section*{Acknowledgements}

YS and YG acknowledge the support from National Key R\&D Program of China grant Nos. 2022YFF0503404, 2020SKA0110402, and the CAS Project for Young Scientists in Basic Research (No. YSBR-092). KCC acknowledges the support the National Science Foundation of China under the grant number 12273121. XLC acknowledges the support of the National Natural Science Foundation of China through Grant Nos. 11473044 and 11973047, and the Chinese Academy of Science grants ZDKYYQ20200008, QYZDJ-SSW-SLH017, XDB 23040100, and XDA15020200. GLL, YL and CLW acknowledges the support from NSFC grant No. U1931210. This work is also supported by science research grants from the China Manned Space Project with Grant Nos. CMS-CSST-2021-B01 and CMS-CSST-2021-A01.

\section*{Data Availability}

 The data that support the findings of this study are available from the corresponding author upon reasonable request.



\bibliographystyle{mnras}
\bibliography{vsfre} 




\appendix
\section{MCMC results for $\beta$}\label{sec:appendix}
\begin{figure}
	\includegraphics[width=\columnwidth]{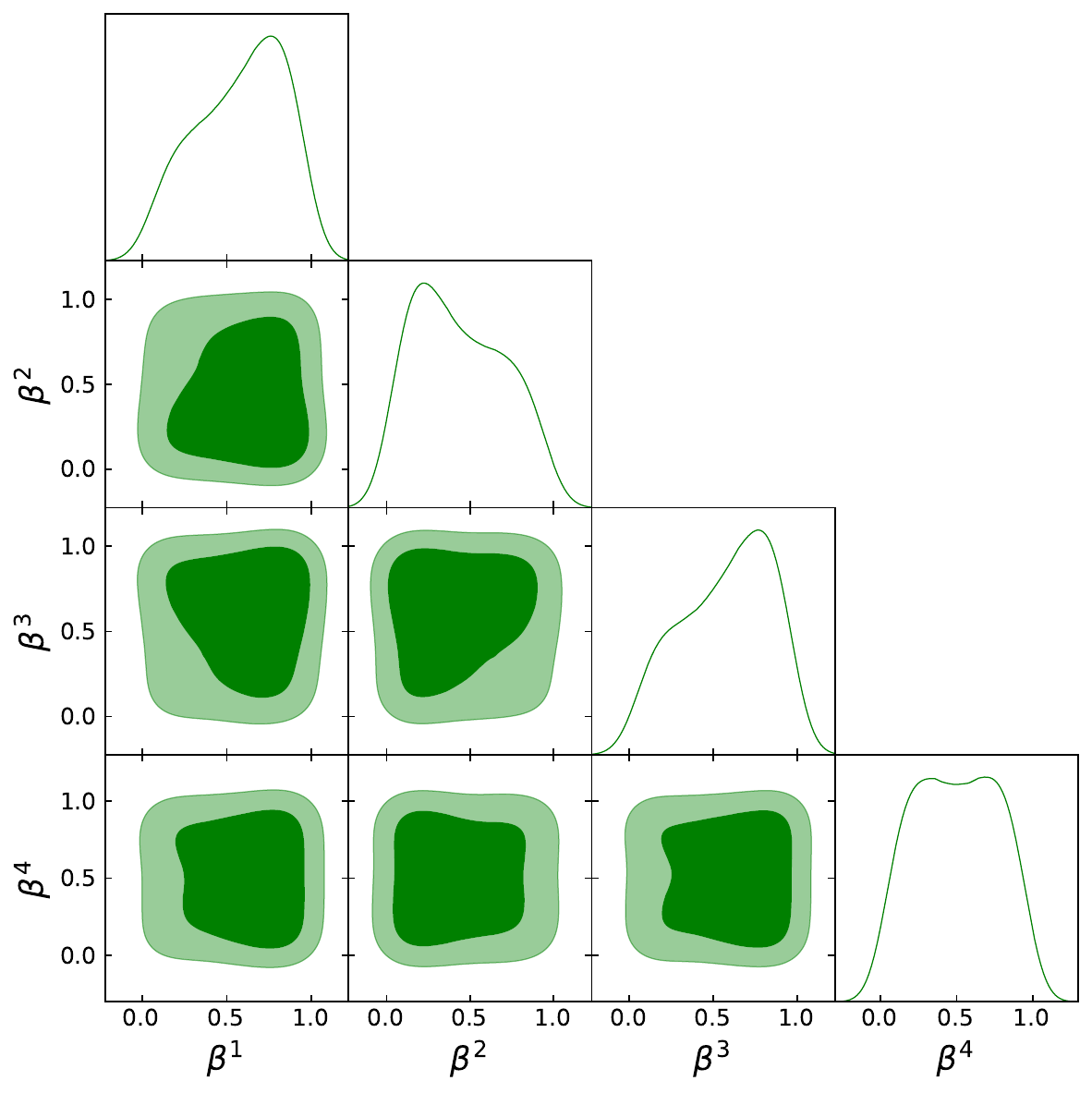}
    \caption{Contour maps of the $\beta^i$ at 68\% and 95\% CL using the VSF mock data at all four redshift bins from $z=0.5$ to 1.1. The 1D PDF (green curve) for each parameter is also shown.}
    \label{fig:mcmcbeta}
\end{figure}

We set $\beta$ in Equation~(\ref{eq18}) as free parameters in different redshift bins, and show the constraint results in green contours and 1D PDFs in Figure {\ref{fig:mcmcbeta}}. We can find that the constraint power of $\beta^i$ is relatively weak, which is due to that the VSF is actually not very sensitive to $\beta$.


\bsp	
\label{lastpage}
\end{document}